# Development of physics-based compositional parameters for predicting the reactivity of amorphous aluminosilicates in alkaline environments

Kai Gong[1,2], Claire E. White[1]

[1]Department of Civil and Environmental Engineering and the Andlinger Center for Energy and the Environment, Princeton University, Princeton, New Jersey 08544, United States

[2]Department of Civil and Environmental Engineering, Rice University, Houston, Texas 77005, United States (current address)

**Abstract**

The reactivity of amorphous aluminosilicates in alkaline environments is important to many applications, including blended Portland cements and alkali-activated materials (AAMs). Here, two physics-based compositional parameters are derived to describe the relative reactivity of CaO-$Al_2O_3$-$SiO_2$ and CaO-MgO-$Al_2O_3$-$SiO_2$ glasses in alkaline environments: (i) a modified average metal oxide dissociation energy (AMODE) parameter developed from molecular dynamics simulations; and (ii) a topology constraint parameter derived from topological constraint theory. Both parameters are seen to generally outperform existing compositional parameters from the literature for a wide range of aluminosilicate glasses. Given that both the modified AMODE and topology constraint parameters can be calculated directly from chemical compositions, this study represents a major step forward in connecting aluminosilicate glass compositions with their reactivity in alkaline environments. The limitations of these compositional parameters, including their inability to capture the impact of Ca versus Mg on glass reactivity, have also been discussed.

## 1   Introduction

Amorphous CaO-$Al_2O_3$-$SiO_2$ (CAS) and CaO-MgO-$Al_2O_3$-$SiO_2$ (CMAS) glassy phases are important constituents of certain supplementary cementitious materials (SCMs) in blended cement systems and high-Ca precursor materials for the synthesis of alkali-activated materials (AAMs). Given the inherent chemical variability of these amorphous phases in SCMs and precursor materials (e.g., ground granulated blast-furnace slag (GGBS), fly ash, and volcanic ash), it is important to understand how their chemical composition influences reactivity in alkaline environments. Numerous studies in the literature have shown that the chemical composition of these SCMs and precursor materials has a large impact on their reactivity in both blended cements and AAMs, as well as the engineering properties of the final cementitious products [1-8], in addition to the impact of activator chemistry for AAMs, curing conditions, and the amorphicity and particle size distribution of the SCMs and AAM precursors [1, 7, 9-12]. A brief literature review on the impact of the different oxide components on the reactivity of CAS and CMAS glassy phases in alkaline environments has been given in our previous investigation [13]. Hence, from an industrial viewpoint, it is important to develop simple methods that are able to predict the relative reactivity of these amorphous phases in alkaline environments.

Many tests have been developed to evaluate the reactivity of SCMs in alkaline conditions, as recently summarized by Snellings and Scrivener [14] and Li *et al.* [15]. Most of these methods [14-16] are based on the cumulative heat release from isothermal conduction calorimetry (ICC) tests, with a higher ICC cumulative heat value indicating higher reactivity. Alternatively, the degree of reaction (and hence reactivity) of these amorphous SCMs or precursor materials can be quantitatively or qualitatively evaluated using other experimental techniques, including X-ray diffraction (XRD) [8, 17-20], thermogravimetric analysis (TGA) [3, 4, 8, 20-22], scanning electron microscopy image analysis (SEM-IA) [3, 4, 19, 23], nuclear magnetic resonance (NMR) [5, 7, 8, 24-27], Fourier transform infrared spectroscopy (FTIR) [12, 28], and more recently, X-ray pair distribution function (PDF) analysis [28].

Alternatively, empirical methods mostly based on chemical composition have also been used to evaluate the reactivity of SCMs and precursor materials in alkaline environments. These compositional parameters, including those used in country-specific standards (e.g., the $(CaO+MgO)/SiO_2$ ratio for the European Standard for slag cement (1994)), have been briefly summarized by Shi *et al.* [29] and Blotevogel *et al.* [9]. Among these parameters, the extent of depolymerization or the average non-bridging oxygen (NBO) species per network former (NBO/T, where T = Si and Al in tetrahedral coordination) is the most commonly used parameter in the glass community to link composition with glass reactivity as well as other glass properties [30]. Several studies have shown direct linkages between NBO/T and aluminosilicate glass reactivity in alkaline environments, with a higher NBO/T value generally resulting in higher reactivity [8, 31]. The NBO/T parameter has also been used in the geochemistry community to evaluate the dissolution rates of silicate minerals [32]. However, it has been shown that the NBO/T parameter is not always a reliable indicator of glass reactivity [8, 33], and silicate mineral dissolution experiments show that the dissolution rates can vary by several orders of magnitude for the same value of NBO/T (e.g., $Mg_2SiO_4$ and $Ca_2SiO_4$ with an NBO/T of 4) [32].

Another approach, topological constraint theory (TCT), has shown that the reactivity of these glassy phases is dependent on the average number of constraints on each atom in the disordered aluminosilicate network, with a higher number of constraints (more rigid structure) leading to generally lower dissolution rates [34, 35]. Recently, we developed two structural descriptors based on force field molecular dynamics (MD) simulations and associated atomic structural representations of CAS and CMAS glasses with a range of chemical compositions [13]. The first structural descriptor we termed the average metal oxide dissociation energy (AMODE), which gives an estimate of the average energy required to break metal-oxygen bonds in the oxide glass, with a higher AMODE value correlated with lower reactivity. The second structural descriptor, the average self-diffusion coefficient (ASDC) of all the atoms in a glass structure at a temperature above melting (e.g., 2000 K), also reflects the ease of metal-oxygen bond breaking, with a higher ASDC value (and hence higher atomic mobility) correlated with higher reactivity. More recently, an average metal–oxygen (M–O) bond strength parameter has been developed based on classical bond valence models and MD simulation results, which exhibited a strong inverse correlation with the reactivity (based on logarithmic silicon dissolution rate data collected at far-from-equilibrium steady-state) of highly complex volcanic glasses (in the compositional space of CaO–MgO–$Al_2O_3$–$SiO_2$–$TiO_2$–FeO–$Fe_2O_3$–$Na_2O$–$K_2O$ (CMASTFNK)) [36]. All three parameters have exhibited superior performance to the NBO/T parameter over a range of C(M)AS [13] and CMASTFNK [36] compositions.

While empirically derived compositional parameters are simple to use, their reliability in predicting glass reactivity is often questionable. In contrast, methods based on experiments (e.g., ICC [14-16], quantitative XRD [8, 17], and NMR [5, 7, 24]) or MD simulations (e.g., TCT [34], AMODE and ASDC [13], and the average M-O bond strength [36]) are generally more reliable. However, these methods require either experimental or computational facilities and associated expertise. As a result, it is desirable to develop physics-based compositional parameters that are directly calculated from chemical compositions of amorphous and/or glassy phases, which combine the benefits of (i) easiness of application and (ii) reliability.

This study investigates the ability of two novel physics-based compositional parameters to predict the relative reactivity of CAS and CMAS glasses in alkaline environments, where the performance of these compositional parameters is benchmarked against other compositional parameters reported in the literature. The first parameter, known as the modified AMODE parameter, is developed based on its structural parameter analog [13]. Specifically, the modified AMODE parameter is based on the MD simulation results of eighteen CAS and CMAS glasses from our previous investigation that covered a wide range of compositions [13]. The second parameter, a topology constraint parameter, is derived from topology constraint theory following the method of calculation described in ref. [37]. The performance of these two physics-based compositional parameters in describing different reactivity data from four different studies [3-5, 8] is then benchmarked against fifteen other compositional parameters summarized in refs. [9, 29]. Further benchmarking is carried out for the modified AMODE and topology constraint parameters against the best-performing existing compositional parameters using data from three recent high-quality investigations [9, 24, 31], which have explored the reactivity of GGBSs and synthetic aluminosilicate glasses in different alkaline environments. From the benchmarking, we show that the two physics-based compositional parameters are generally more reliable than existing compositional parameters, including the commonly used basicity coefficient and extent of depolymerization.

## 2  Glass Compositions

Previously, we have employed force field MD simulations to generate detailed structural representations for ten CMAS glasses and eight CAS glasses with a wide range of chemical compositions relevant to GGBS and fly ash [13]. The chemical compositions of the eighteen CMAS and CAS glasses are summarized in Table 1, along with the average coordination numbers (CNs) for Ca, Mg, Si, and Al atoms in each glass, obtained from the analysis of MD simulation results [13]. The CMAS glasses in Groups A-C have the same $CaO-MgO-Al_2O_3-SiO_2$ compositions as the ten GGBSs from refs. [3-5], which are predominately amorphous as evidenced by the XRD data in the relevant studies. In ref. [5], the impact of MgO content on the reactivity of Group A GGBSs during $Na_2CO_3$ activation was investigated using ICC. GGBSs in Groups B and C were studied in refs. [3] and [4], respectively, where the impact of MgO and $Al_2O_3$ content on GGBS reactivity during NaOH and $Na_2SiO_3$ activation was explored using TGA. However, as shown in our previous study, the main oxide compositions in each group of GGBSs are interconnected and, hence, the observed difference in their reactivity in alkaline environments should not simply be attributed to their compositional difference in one oxide component (e.g., MgO or $Al_2O_3$). The MD simulation results show that the average CNs of the Ca, Mg, Si, and Al atoms in the ten GGBSs do not vary significantly: 4 for Si, 4.02-4.04 for Al, 4.99-5.19 for Mg, and 6.72-6.83 for Ca atoms [13]. The CNs were calculated from the last 500 ps of MD trajectories (averaged over 500 configurational snapshots) during the *NVT* equilibration step at 300 K with cutoff distances of 2.2, 2.5, 2.9, and 3.2 Å for Si-O, Al-O, Mg-O, and Ca-O pairs, respectively, as discussed in our previous investigation [13].

The CAS compositions in Group D are included to represent the eight synthetic CAS glasses from another study [8], which cover a wider compositional range than the CMAS glasses in Groups A-C, as shown in Table 1. According to the XRD data in ref. [8] these synthetic CAS glasses are also predominantly amorphous, and their reactivity has been evaluated in a blended mixture of NaOH, $Ca(OH)_2$, and limestone using quantitative XRD [8]. The D8 glass was included to represent a GGBS composition without MgO, while the D1-3 and D5-7 compositions were designed to assess the impact of $Al_2O_3$ content on CAS reactivity in the Si-rich fly ash region (e.g., class F) and Ca-rich fly ash region (e.g., class C), respectively. D2, D4, and D6 compositions were used to study the impact of Ca/Si ratio on CAS glass reactivity at fixed $Al_2O_3$ content (i.e., ~26 wt. %). The MD simulation results (Table 1) show that the Si atoms are all four-fold coordinated (similar to the CMAS glasses in Groups A-C), while the CNs of Al and Ca atoms exhibit a slightly larger range of values than those of the CMAS glasses in Groups A-C: 4.02-4.11 for Al and 6.49-7.19 for Ca atoms. A detailed comparison of the MD simulation results with the available experimental data has been carried out in ref. [13] (including the nearest interatomic distances and CNs), which shows that the MD-generated structural representations have captured the key structural features of these CMAS and CAS glasses.

Table 1. The number of atoms in each simulation box (corresponding to the oxide composition of each GGBS or synthetic CAS glass from refs. [3-5, 8]) used in the force field MD simulations conducted in ref. [13]. The average coordination numbers (CNs) for the Si, Al, Mg, and Ca atoms in each glass, obtained from the MD simulations [13], are also given in the table. The CN values in the table are averages based on three independent MD trajectories, with one standard deviation given in the brackets. Finally, the average CN for each type of atom (i.e., Si, Al, Mg, and Ca) based on all the eighteen glasses is also given at the bottom of the table.

| Glass ID # | Number of atoms in the simulation box | | | | | | Average coordination number | | | |
|---|---|---|---|---|---|---|---|---|---|---|
| | Ca | Mg | Si | Al | O | Total | Si | Al | Mg | Ca |
| A1_1Mg | 394 | 15 | 271 | 148 | 1173 | 2001 | 4.00 (0.00) | 4.03 (0.01) | 4.99 (0.07) | 6.75 (0.02) |
| A2_5Mg | 375 | 65 | 267 | 130 | 1169 | 2006 | 4.00 (0.00) | 4.02 (0.01) | 5.07 (0.10) | 6.76 (0.05) |
| A3_7Mg | 355 | 78 | 290 | 106 | 1172 | 2001 | 4.00 (0.00) | 4.03 (0.01) | 5.17 (0.05) | 6.75 (0.01) |
| A4_14Mg | 286 | 169 | 294 | 82 | 1166 | 1997 | 4.00 (0.00) | 4.04 (0.02) | 5.13 (0.02) | 6.80 (0.01) |
| B1_8Mg | 307 | 93 | 306 | 114 | 1183 | 2003 | 4.00 (0.00) | 4.04 (0.03) | 5.09 (0.06) | 6.74 (0.02) |
| B2_11Mg | 296 | 126 | 296 | 108 | 1176 | 2002 | 4.00 (0.00) | 4.03 (0.01) | 5.15 (0.09) | 6.78 (0.06) |
| B3_13Mg | 282 | 156 | 287 | 104 | 1168 | 1997 | 4.00 (0.00) | 4.03 (0.01) | 5.19 (0.05) | 6.83 (0.02) |
| C1_7Al | 334 | 86 | 332 | 66 | 1183 | 2001 | 4.00 (0.00) | 4.02 (0.01) | 5.04 (0.06) | 6.73 (0.01) |
| C2_14Al | 305 | 78 | 302 | 132 | 1185 | 2002 | 4.00 (0.00) | 4.02 (0.03) | 5.04 (0.09) | 6.72 (0.02) |
| C3_17Al | 294 | 75 | 292 | 154 | 1184 | 1999 | 4.00 (0.00) | 4.03 (0.001) | 5.14 (0.12) | 6.77 (0.03) |

| | | | | | | | 4.00 | 4.04 | | |
|---|---|---|---|---|---|---|---|---|---|---|
| D1 | 34 | 0 | 532 | 134 | 1299 | 1999 | (0.00) | (0.005) | | 6.69 (0.01) |
| D2 | 34 | 0 | 469 | 210 | 1287 | 2000 | 4.00 (0.00) | 4.11 (0.01) | | 7.00 (0.09) |
| D3 | 31 | 0 | 412 | 280 | 1275 | 1998 | 4.00 (0.00) | 4.09 (0.01) | | 6.49 (0.04) |
| D4 | 104 | 0 | 414 | 220 | 1262 | 2000 | 4.00 (0.00) | 4.06 (0.02) | | 6.77 (0.08) |
| D5 | 163 | 0 | 442 | 140 | 1257 | 2002 | 4.00 (0.00) | 4.07 (0.03) | | 7.00 (0.07) |
| D6 | 186 | 0 | 358 | 222 | 1235 | 2001 | 4.00 (0.00) | 4.07 (0.01) | | 7.19 (0.03) |
| D7 | 185 | 0 | 286 | 308 | 1219 | 1998 | 4.00 (0.00) | 4.06 (0.01) | | 7.11 (0.02) |
| D8 | 417 | 0 | 272 | 140 | 1171 | 2000 | 4.00 (0.00) | 4.02 (0.003) | | 6.82 (0.02) |
| | | | | Average CN based on all glasses: | | | **4.00** | **4.05** | **5.10** | **6.81** |

## 3 Results and Discussion

### 3.1 Average metal oxide dissociation energy

Based on the MD simulation results of the eighteen CMAS and CAS glasses reported in Table 1, two structural descriptors, i.e., AMODE and ASDC at 2000 K, have been developed in our previous investigation [13], which were shown to give an accurate description of the different reactivity data collected on these CMAS and CAS glass compositions from the four literature investigations [3-5, 8]. Here, we modify the AMODE parameter from being based on the structure of the glass (i.e., a structural descriptor) to solely relying on the chemical composition of the glass (i.e., a compositional parameter).

The AMODE parameter, which gives an estimation of the average energy required to break (or dissolve) metal-oxygen bonds in the glass, is defined using Equation (1).

$$AMODE = \frac{\sum N_M \cdot CN_M \cdot BS_{M-O}}{\sum N_i} \quad (1)$$

where $N_M$ is the number of each type of metal cation ($M$ = Ca, Mg, Al, or Si) in the glass, $CN_M$ and $BS_{M-O}$ are the average CN and metal-oxygen single bond strength (BS) for cation $M$, respectively. As previously explained in ref. [13], the average metal-oxygen single BS values are obtained or estimated from the literature data [38]: (i) 32, 37 and 106 kcal for Ca-O, Mg-O, and Si-O single bonds in six-, six-, and four-fold coordination, respectively; and (ii) 90, 75, and 60 kcal for Al-O single bonds in four-, five-, and six-fold coordination, respectively. In contrast, the $CN_M$ values are calculated from the MD simulation results for each glass composition (as shown in Table 1). Although there are several limitations with the estimation of the AMODE parameter, as extensively discussed in ref. [13], we previously showed in the same reference that this AMODE parameter gives an accurate description of different reactivity data (including ICC cumulative heat, TGA bound water content, and degree of reaction from quantitative XRD analysis) [3-5, 8] collected on the eighteen GGBSs and CAS glasses (reported in Table 1) exposed to different alkaline environments. The $R^2$ values for linear regressions between the AMODE parameter and the different reactivity data were found to be generally higher than 0.90 [13].

Alternatively, the $CN_M$ values can also be determined experimentally (e.g., using NMR) to estimate the AMODE parameter. However, the requirement of experimental measurements or MD simulations renders the widespread application of Equation 1 difficult. To increase its applicability, here we modify the AMODE parameter so that it only depends on the chemical composition of the raw CMAS and CAS glass. From Equation 1, we see that the only inputs taken from the MD simulations are the CNs of the different metal cations (i.e., Ca, Mg, Al, and Si), while the BS values are obtained or estimated from the literature. Given that the CNs of Ca, Mg, Al, and Si atoms are relatively similar from one CMAS/CAS composition to another (with differences equal to or smaller than ~11%, 4%, 3%, and 0% for Ca, Mg, Al, and Si atoms, respectively, according to the MD simulation results in Table 1), it is reasonable to simplify Equation 1 by using the average CN values. Here, the average CN values based on the CN results for all the eighteen CMAS/CAS glass compositions in Table 1 are used: i.e., 6.81, 5.10, 4.05, and 4, for Ca, Mg, Al, and Si atoms, respectively. This modified AMODE parameter is given by Equation 2, which depends only on the chemical composition of the CMAS/CAS glasses.

$$Modified\ AMODE = \frac{218N_{Ca} + 189N_{Mg} + 360N_{Al} + 424N_{Si}}{(N_{Ca} + N_{Mg} + N_{Al} + N_{Si})} \qquad (2)$$

where $N_{Ca}$, $N_{Mg}$, $N_{Al}$, and $N_{Si}$ are the molar concentrations of Ca, Mg, Al, and Si atoms in the glass, respectively.

To evaluate the validity of this simplification, the modified AMODE parameter for all the eighteen glasses (in Table 1) has been compared with the original AMODE parameter in Figure 1, where it is seen that the modified AMODE parameter is linearly correlated with (an $R^2$ value of 0.999) and almost the same as the original AMODE parameter for all the compositions considered. The largest deviation is less than 1%, which shows that the simplification (i.e., using the same average CN values for all glasses for each type of atom) is reasonable.

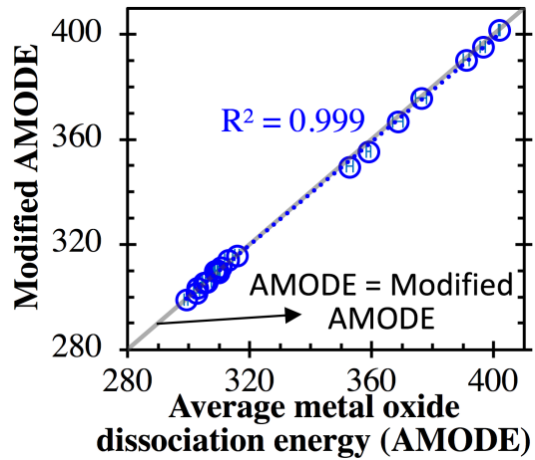

Figure 1. Comparison of (i) the average metal oxide dissociation energy (AMODE) determined using Equation 1 and (ii) the modified AMODE parameter calculated using Equation 2, obtained from the results of the eighteen CMAS/CAS glasses listed in Table 1.

Next, the performance of the modified AMODE parameter has been evaluated by examining its correlation with the different reactivity data collected on these CMAS and CAS glasses under different alkaline environments [3-5, 8], as illustrated in Figure 2. We note that the reactivity data in Figure 2 have been normalized by the surface area of each GGBS or CAS glass using a similar approach adopted in ref. [13], as the surface area has been shown to be linearly and positively correlated with reactivity (a higher surface area leads to a higher reactivity) [9]. According to its definition, a higher modified AMODE value is an indication of a stronger average metal-oxygen bond for a given phase that requires, on average, more energy to break, and as a result, the phase should have a higher reactivity (i.e., a greater extent of reaction at a given time). Therefore, the modified AMODE parameter is seen to be positively correlated to the time to reach the first ICC reaction peak (a shorter time to reach the ICC peak means higher reactivity; Figure 2a) and inversely correlated with TGA bound water content (a reflection of the degree of reaction; Figure 2b-c) and the degree of reaction from quantitative XRD data (Figure 2d). These trends are consistent with those of the original AMODE parameter, as reported in ref. [13], where the same sets of experimental data were used for comparison. A logarithmic scale of ICC time is used for the x-axis in Figure 2a (as opposed to a linear scale adopted for other reactivity data in Figure 2b-d) because the degree of reaction (or ICC cumulative heat curve) is approximately a logarithmic function with time [13]. Given that the modified AMODE parameter is almost linearly correlated with the original AMODE parameter (as seen in Figure 1), it is not surprising that the level of agreement achieved in Figure 2 when compared with experimental data, as evaluated by the $R^2$ values (0.86-0.97), is comparable to that of the original AMODE parameter in our previous investigation ($R^2$ = 0.93-0.97) [13].

Furthermore, we have also used this modified AMODE parameter to correlate with other reactivity data available in refs. [3-5, 8], including the degree of reaction data from NMR and/or thermodynamic modeling, compressive strength data, and TGA bound water data collected on NaOH-activated GGBSs. The results are given in Figure S1 of the Supplementary Material, which shows that the level of agreement (as evaluated using $R^2$ values for linear fits) is comparable to that achieved in Figure 2 for each group of glass in Table 1. This is encouraging given that the modified AMODE parameter can be calculated directly from the C(M)AS glass composition without the need for time-consuming simulations or experiments.

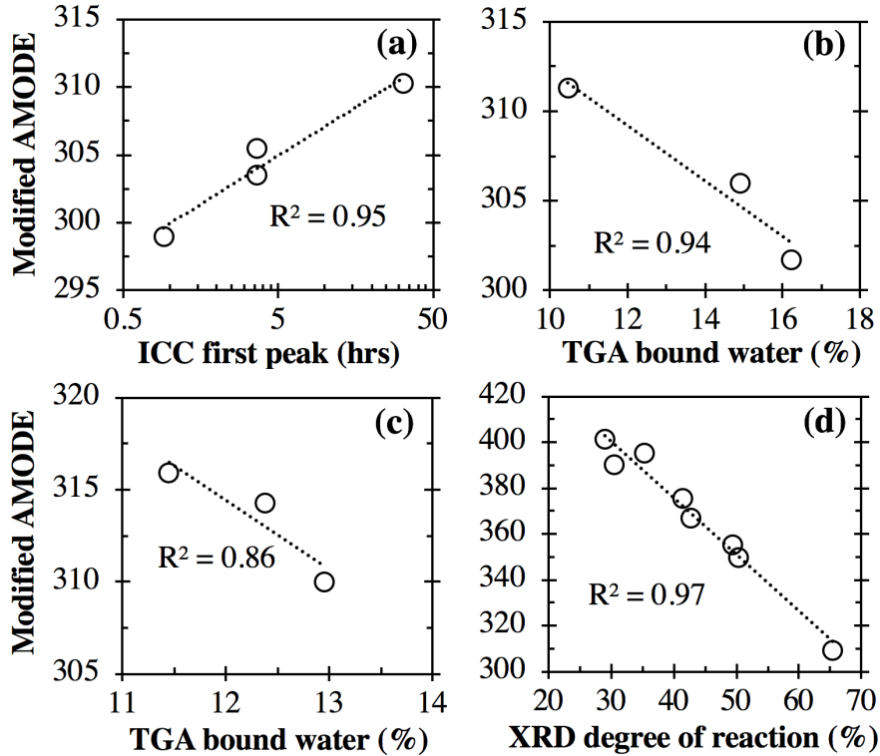

Figure 2. Comparison of the modified average metal oxide dissociation energy (AMODE) parameter (calculated using Equation 2) of the CMAS and CAS glasses with the corresponding reactivity data for (a) Group A [5], (b) Group B [3], (c) Group C [4], and (d) Group D [8] glasses. The isothermal conduction calorimetry (ICC) data (time to reach the first ICC peak) in (a) were obtained from ref. [5] based on $Na_2CO_3$-activated GGBS with Group A chemical composition. The thermogravimetric analysis (TGA) bound water content data in (b) and (c) were collected on 28-day $Na_2SiO_3$-activated GGBS with Group B and C composition, obtained from refs. [3] and [4], respectively. The degree of reaction data in (d) was obtained from ref. [8] based on quantitative X-ray diffraction (XRD) analysis of synthetic CAS glasses (Group D composition) in a blended mixture of NaOH, $Ca(OH)_2$, and $CaCO_3$ and cured for 180 days. A linear fit between the modified AMODE parameter and the reactivity data (dotted line), along with the $R^2$ value (goodness of fit), are given in each plot.

## 3.2 Topology constraint $n_r$

TCT has emerged as a promising tool linking glass structure and material properties, and has recently been employed to understand aluminosilicate glass dissolution in alkaline environments [34, 35]. TCT works by using mechanical trusses to simplify the complexity of atomic structures, where the number of constraints per atom is calculated to estimate the overall rigidity of the truss. Two types of constraints are usually utilized, i.e., radial bond-stretching (BS) constraint (e.g., Si-O stretching) and angular bond-bending (BB) constraint (e.g., O-Si-O bending). According to ref. [37], for calcium aluminosilicate glasses, each Si and Al atom in four-fold coordination has five angular BB constraints (O-Si-O or O-Al-O). In addition, each O atom has two BS constraints (Si-O, Al-O, or Ca-NBO) and one BB constraint (Si-O-Si, Si-O-Al, Al-O-Al, Si-NBO-Ca, or Al-NBO-Ca), where NBO refers to non-bridging oxygen (defined as an oxygen atom only connected to one network former, i.e., Si and Al atoms in four-fold coordination). Each excess Ca atom beyond those required for charge-balancing the negatively charged four-fold Al atoms ($[Al(O_{1/2})_4]^{-1}$) creates two BS constraints ($Ca^{2+}$-NBO), according to ref. [37]. Given that Mg is generally considered as a network modifier, similar to Ca in aluminosilicate glasses [39], here we will treat the Mg atoms as the Ca atoms for estimation of topology constraints. For Al atoms in five-fold coordination, each atom has seven BB constraints. Finally, we will use the same average CN for the Al atom in all the glasses (i.e., 4.05 in Table 1) and assume that five-fold coordination is the only high-coordination state for the Al atom (i.e., no six-fold coordinated Al atoms are present). This assumption is based on the fact that both experiments and simulations have shown that the dominant high-coordination state in aluminosilicate glasses is five-fold [7, 39]. Hence, the average number of constraints per atom ($n_r$) in the CMAS and CAS glasses can be estimated using equation (3).

$$n_r = \frac{3N_O + 5N_{Si} + 4.15N_{Al} + 2N_{Ca} + 2N_{Mg}}{N_O + N_{Si} + N_{Al} + N_{Ca} + N_{Mg}} \tag{3}$$

where $N_O$, $N_{Si}$, $N_{Al}$, $N_{Ca}$, and $N_{Mg}$ are the molar fractions of O, Si, Al, Ca, and Mg atoms in each glass. Given that $N_O = 2N_{Si} + 1.5N_{Al} + N_{Ca} + N_{Mg}$ (based on the number of oxygen atoms associated with each oxide, e.g., $SiO_2$, $Al_2O_3$, CaO and MgO), Equation 3 can be simplified to give Equation 4.

$$n_r = \frac{11N_{Si} + 8.65N_{Al} + 5N_{Ca} + 5N_{Mg}}{3N_{Si} + 2.5N_{Al} + 2N_{Ca} + 2N_{Mg}} \tag{4}$$

To evaluate the performance of the topology constraint parameter $n_r$ (in Equation 4), we have compared it with the different reactivity data in Figure 3 (the reactivity data has been normalized following ref. [13]), similar to the modified AMODE parameter comparison performed in Figure 2. The data show that this parameter, $n_r$, is (i) linearly and positively correlated with the logarithmic time to reach the first ICC reaction peak for the $Na_2CO_3$-activated GGBSs (with an $R^2$ value of 0.90, Figure 3a); (ii) inversely correlated with the TGA bound water content in the $Na_2SiO_3$-activated GGBSs (with $R^2$ values of 0.95 and 0.85 in Figure 3b and 3c, respectively); and (iii) inversely correlated with the degree of reaction data from quantitative XRD analysis of the synthetic CAS glasses in a blended mixture of NaOH, $Ca(OH)_2$ and $CaCO_3$ (with an $R^2$ value of 0.96, Figure 3d). These general trends are similar to those obtained for the modified AMODE parameter in Figure 2, and the degrees of correlation (as evaluated by the $R^2$ values achieved for linear regressions) are also comparable between the two parameters. These similar trends are expected because a higher $n_r$ value means the glass structure is more rigid and harder to break or dissolve. As a result, a glass with a higher $n_r$ value should also exhibit a higher modified AMODE value. Despite this expectation, it is still interesting to see in Figure 4 that the modified AMODE parameter is almost linearly correlated with the topology constraint parameter $n_r$, with an $R^2$ value of 0.998 for linear regression, given that they are derived based on different concepts.

Finally, we have assessed the degree of correlation between this topology constraint parameter $n_r$ and other reactivity data available in refs. [3-5, 8], specifically, the degree of reaction from NMR and/or thermodynamic modeling, compressive strength data, and TGA bound water data collected on NaOH-activated GGBSs. The results are presented in Figure S2 of the Supplementary Material, and the level of agreement (as evaluated using $R^2$ values for linear fits) is comparable to that achieved in Figure 3 for each group of glass studied here.

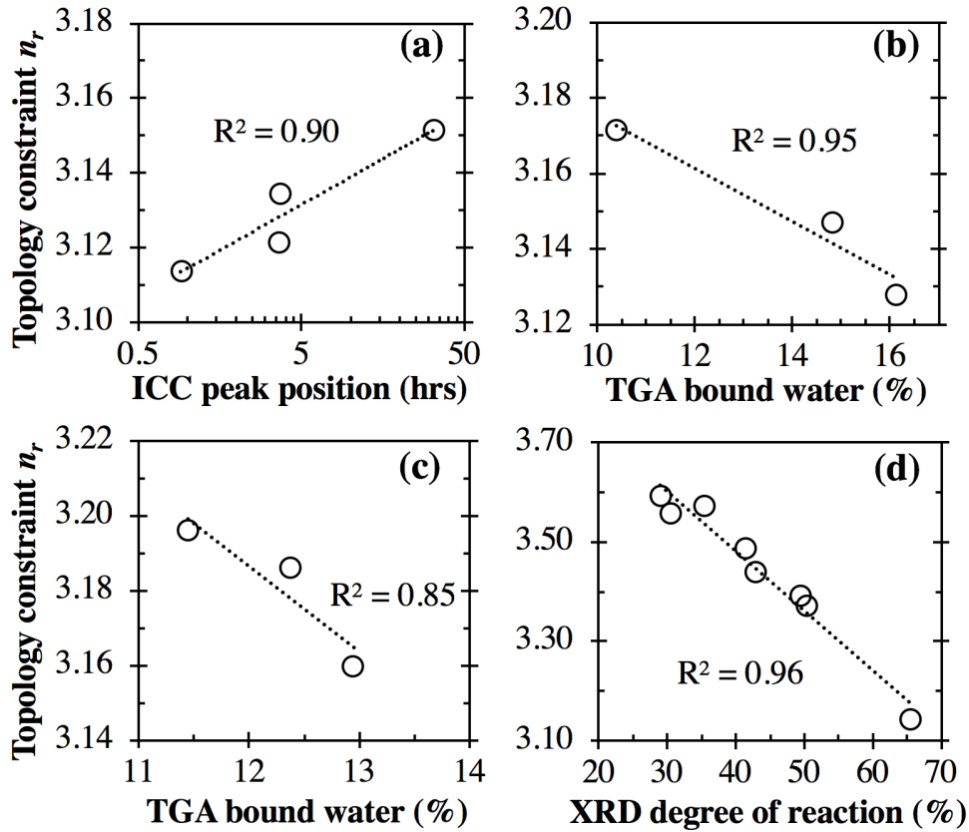

Figure 3. Comparison of the topology constraint parameter $n_r$ (estimated using Equation 4) of the CMAS and CAS glasses with the corresponding reactivity data collected for (a) Group A [5], (b) Group B [3], (c) Group C [4], and (d) Group D [8] glasses. Details about these experimental data [3-5, 8] have been given in the caption of Figure 2. A linear fit between the $n_r$ parameter and the reactivity data (dotted line) is given in each plot, with the $R^2$ value (goodness of fit) also given.

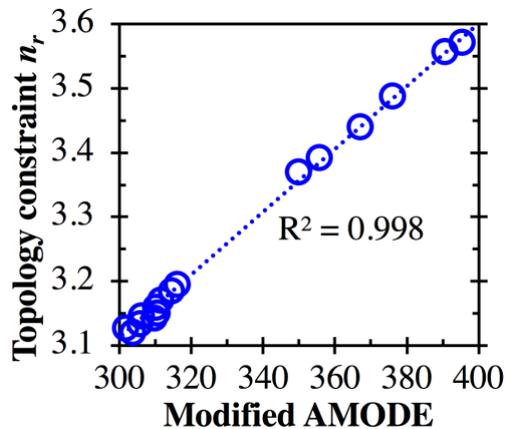

Figure 4. Comparison of the modified AMODE parameter calculated using Equation 2 with the topology constraint parameter, $n_r$, calculated using Equation 4 for all the CMAS and CAS glass compositions in Table 1. A linear fit between the $n_r$ parameter and the modified AMODE parameter is depicted by the blue dotted line, with the $R^2$ value (goodness of fit) also given.

## 3.3 Comparison with other commonly used compositional parameters

Many compositional parameters (outlined in Table 2) have been used in the literature and country-specific standards as quality criteria to gauge the reactivity of SCMs in blended cements or precursor materials in AAMs, as has been summarized in refs. [29, 40]. The majority of these mostly empirically-derived parameters consist of a quotient, where the CaO content is often found in the numerator and the $SiO_2$ content in the denominator, a reflection of the disparate behavior of Ca and Si on glass reactivity: Ca atoms are network modifiers with weaker Ca-O bonds (~32 kJ/mol for a single bond [38]) that are relatively easy to break, whereas Si atoms are network formers with stronger Si-O bonds (~106 kJ/mol for a single bond [38]). A major difference between the parameters in Table 2 centers on the positions of MgO and $Al_2O_3$ in the numerator and denominator along with their coefficients, and contributions from minor oxide components (e.g., $TiO_2$ and MnO in parameters K8-K11 of Table 2).

In most cases, MgO is placed on the same side as CaO, since Mg atoms are generally considered network modifiers (similar to Ca atoms), although their potential role as network formers has also been postulated [41]. It is interesting to note that $Al_2O_3$ is positioned mostly on the side of network modifiers (e.g., Ca atoms) in these parameters (i.e., K3-K7 and K9-K11), although Al atoms in compositions reminiscent of GGBS and fly ash should act mostly as network formers (similar to Si atoms) because Al atoms are mostly in four-fold coordination for these compositions according to both experiments and simulations [8, 13, 39, 42], with an average Al-O single bond strength of ~90 kJ/mol (79-101 kJ/mol, [38]). This behavior of Al atoms as network formers has been captured by the parameters K13 and K14, where $Al_2O_3$ is positioned on the same side of the quotient as $SiO_2$. On the other hand, higher coordination Al atoms (five- and six-fold) are generally considered network modifiers [42], and the average Al-O single bond strength for Al in six-fold coordination is estimated to be ~60 kJ/mol (53-67 kJ/mol, [38]), which is closer to the single bond strength of Mg-O (~37 kJ/mol, [38]) and Ca-O bonds (~32 kJ/mol, [38]) than that of the Si-O bond (~106 kJ/mol, [38]). According to both experiments and simulations [8, 13, 43], the dominant high coordination state for Al atoms in aluminosilicate glasses is five-fold (as opposed to six-fold), and the potential dual role (network modifier and former) of five-fold Al atoms in aluminosilicate glasses has been suggested [42, 44]. Parameter K12 partially reflects this potential dual role of Al (network modifier and former), as seen in Table 2, where $Al_2O_3$ has been included on both sides of the quotient.

Finally, the parameter K15 in Table 7, the so-called extent of depolymerization (NBO/T), has been widely used as a reactivity indicator in the cement and glass literature [8, 31, 45]. This parameter assumes that all the Al atoms are network formers in four-fold coordination, in agreement with the dominant coordination state for Al atoms in CMAS and CAS glasses [8, 13, 43]. A distinction between Al and Si network formers is that the four-fold Al atom is negatively charged ($[Al(O_{1/2})_4]^{-1}$ vs $[Si(O_{1/2})_4]^0$), and hence requires positively charged cations (e.g., $Ca^{2+}$ and $Mg^{2+}$) for charge-balancing, with the surplus alkaline earth modifiers each creating two NBOs (network modification effect). In this parameter (K15), minor oxide components that exist in the blast furnace slags in ref. [3-5] have also been considered, with $Na^+$, $K^+$ and $Fe^{2+}$ treated as network modifiers while $Mn^{2+}$ and $Ti^{4+}$ are considered as network formers. Some of these minor oxide components (MnO, FeO, and $TiO_2$) have been considered in parameters K8-K11, where they are all on the side of $SiO_2$ (treated as network formers). However, due to their small quantities, the impact of these minor oxide components on these parameters is rather limited, as illustrated in Figure S3 of the Supplementary Material.

Table 2. A summary of compositional parameters that have been used in the literature and country-specific standards as quality criteria to gauge the reactivity of SCMs in blended cements or precursor materials in AAMs [29, 40]. All the parameters are calculated using weight percentage, except for K15, which is based on molar percentage.

| ID # | Compositional parameters | Notes |
|---|---|---|
| K1 | $K1 = 100 - SiO_2$ | |
| K2 | $K2 = (CaO + MgO)/SiO_2$ | European Standard for slag cement (1994) |
| K3 | $K3 = (CaO + MgO + Al_2O_3)/SiO_2$ | A quality criterion for blast furnace slag in many countries, including Europe and Canada |

| K4  | $K4 = (CaO + MgO + Al_2O_3 - 10)/SiO_2$ | |
|-----|------|------|
| K5  | $K5 = (CaO + 1.4MgO + 0.6Al_2O_3)/SiO_2$ | |
| K6  | $K6 = CaO + 0.5MgO + Al_2O_3 - 2SiO_2$ | |
| K7  | $K7 = (6CaO + 3Al_2O_3)/(7SiO_2 + 4MgO)$ | Based on the 28-day strength of an MgO-rich slag |
| K8  | $K8 = (CaO + 0.5MgO + CaS)/(SiO_2 + MnO)$ | |
| K9  | $K9 = (CaO + 0.5MgO + Al_2O_3)/(SiO_2 + FeO + (MnO)^2)$ | |
| K10 | $K10 = (CaO + MgO + Al_2O_3 + BaO)/(SiO_2 + MnO)$ | |
| K11 | $K11 = (CaO + MgO + Al_2O_3)/(SiO_2 + MnO + TiO_2)$ | A quality criterion for blast furnace slag in a Chinese standard |
| K12 | $K12 = (CaO + MgO + 0.3Al_2O_3)/(SiO_2 + 0.7Al_2O_3)$ | Proposed based on 28-day strength; German Standard for Hochofenzement (1932) |
| K13 | $K13 = (CaO + MgO)/(SiO_2 + 0.5Al_2O_3)$ | Proposed based on 28-day strength |
| K14 | $K14 = (CaO + MgO)/(SiO_2 + Al_2O_3)$ | Basicity coefficient; German Standard for Eisenportlandzement (1909) |
| K15 | $K15 = 2(CaO + MgO + Na_2O + K_2O + FeO - Al_2O_3)/(SiO_2 + 2 \times Al_2O_3 + MnO + TiO_2)$ | Extent of depolymerization (i.e., NBO/T, average non-bridging oxygen (NBO) per network former T (T= Si, Al, Mn and Ti) |

Figure 5 shows how some of the compositional parameters (i.e., K3, K5, K13, and K15) in Table 2 correlate with the different reactivity data (already presented in Section 3.1) for Groups A-D glass compositions [3-5, 8], with the $R^2$ values also given in the figure. A complete overview of $R^2$ values achieved for all the compositional parameters in Table 2 is given in Table 3 and Figure S4 of the Supplementary Material. In addition to $R^2$ values, it is equally important to check if the parameters can correctly predict the expected trend: positively or negatively correlated. For the K1-15 parameters listed in Table 2, it is expected that a higher value gives a higher reactivity because all these parameters increase with increasing CaO content and decreasing $SiO_2$ content. Hence, for the ICC data in Figure 5a, an appropriate parameter should not only give a high $R^2$ value but also be inversely correlated with the ICC first peak time (as denoted by "–" in Table 3). Based on this criterion, Table 3 shows that only K14 and K15 have comparable performance to the modified AMODE and topology constraint $n_r$ parameters for the Group A data. K12 and K13 also correctly capture the inverse correlation, but their corresponding $R^2$ values are too low (0.02-0.08). In contrast, all the other parameters (K1-K11 in Table 3) do not possess the expected inverse correlation.

For the TGA bound water content in Figure 5b-c, a good compositional parameter from Table 2 is expected to give a positive correlation (as denoted by "+" in Table 3). It is seen from Figure 5b and Table 3 that all the K parameters (except K7 and K9) give the expected positive correlation for CMAS glasses in Group B along with relatively high $R^2$ values (> 0.9). However, for the CMAS glasses in Group C (Figure 5c), only parameters K12-K15 exhibit a positive correlation along with reasonably high $R^2$ values (0.82-0.85). For the XRD degree of reaction data collected on the synthetic CAS glasses (Group D), we see in Figure 5d and Table 3 that all the parameters (K1-K15) give the expected positive correlation (a higher value indicates a more extensive reaction and hence a higher reactivity), as well as reasonably high $R^2$ values (> 0.78). It is somewhat surprising to see that most of these parameters, which perform poorly for the Groups A-C CMAS glass data in Figure 5a-c, give $R^2$ values above 0.8 for the CAS glass data in Figure 5d, including some of the simplest parameters (e.g., K1-K2). However, given that the CAS glasses in Group D span a much wider compositional range than the CMAS glasses in Groups A-C (see Table 1), the ability for even the simplest parameters to capture the global trend (e.g., K1 and K2 give higher $R^2$ values than K15, the commonly used extent of depolymerization parameter) suggests that the reactivity of these CAS glasses is largely controlled by their $CaO/SiO_2$ molar ratios (which range from ~0.06 to ~1.53). For mild compositional variations seen in the CMAS glasses ($CaO/SiO_2$ molar ratios in the range of ~0.97-1.45, ~0.98-1.00, and ~1.01-1.01 for Groups A-C, respectively), the impact of Mg and Al atoms on the reactivity becomes more relevant, which explains the generally poor performance of K1-K11 for Groups A-C data.

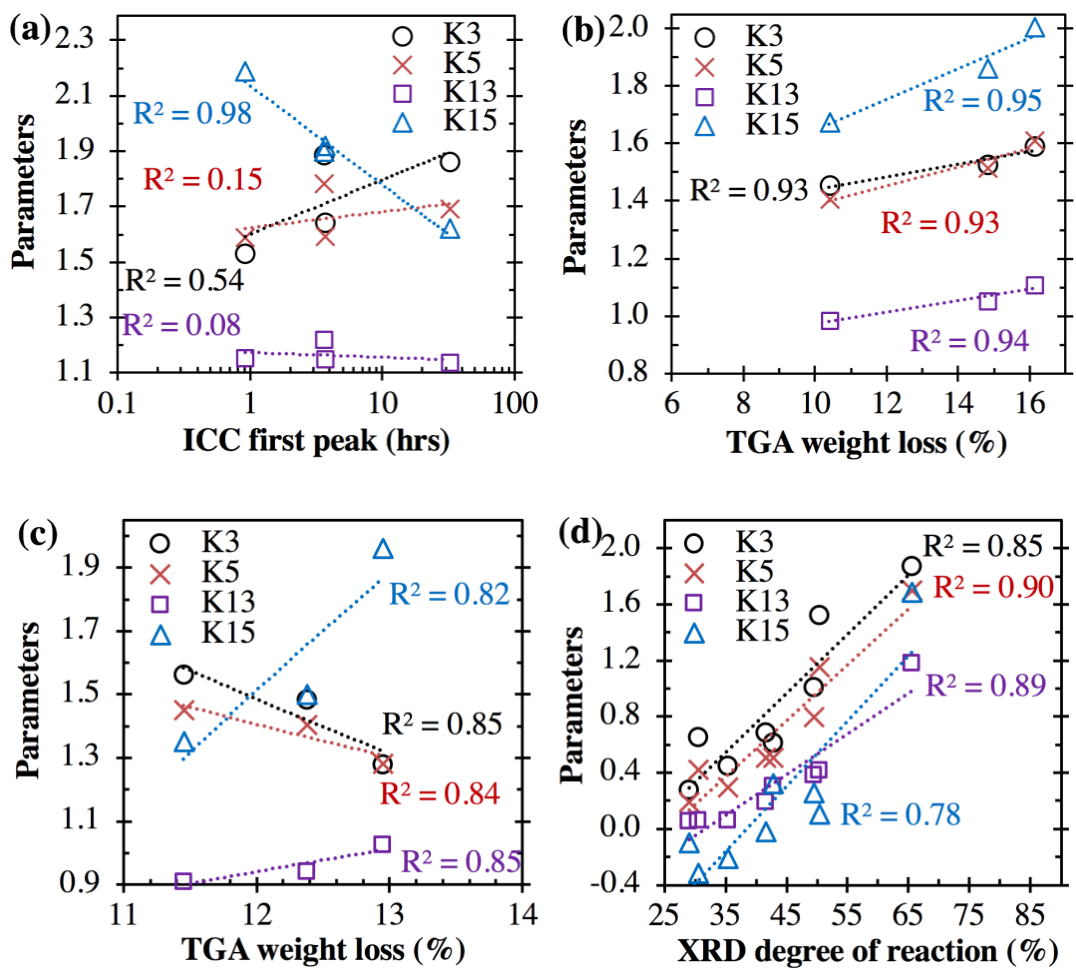

Figure 5. Comparison of reactivity data with representative compositional parameters (K3, K5, K13, and K5 in Table 2) for aluminosilicate glasses in (a) Group A, (b) Group B, (c) Group C, and (d) Group D. Details about these experimental data [3-5, 8] have been given in the caption of Figure 2 and the K compositional parameters are defined in Table 2. Similar plots for the other 11 K parameters in Table 2 are given in Figure S4 of the Supplementary Material.

Table 3. A summary of $R^2$ values achieved between (i) the experimental data in Figure 5 and Figure S4 of the Supplementary Material and (ii) the different K compositional parameters in Table 2 using linear regression. The "+" sign denotes a positive correlation, while the "–" sign denotes an inverse correlation, with the shaded (unshaded) signs indicating consistency (inconsistency) between the predicted and expected trends. $R^2$ values higher than 0.80, indicating good performance, are given in bold. An average $R^2$ value, as defined by Equation 5, is introduced to evaluate the overall performance of these parameters across the four groups of reactivity data. The corresponding $R^2$ values achieved for the modified AMODE and topology constraint $n_r$ parameters are also given for comparison.

| Param. | Group A | | Group B | | Group C | | Group D | | Average $R^2$ value |
| --- | --- | --- | --- | --- | --- | --- | --- | --- | --- |
| | ICC data | Positive or inverse | TGA bound water | Positive or inverse | TGA bound water | Positive or inverse | XRD degree of reaction | Positive or inverse | |
| K1 | 0.67 | + | **0.97** | + | 0.81 | – | **0.82** | + | 0.08 |
| K2 | 0.67 | + | **0.96** | + | 0.82 | – | **0.85** | + | 0.08 |
| K3 | 0.54 | + | **0.93** | + | 0.85 | – | **0.85** | + | 0.10 |
| K4 | 0.52 | + | **0.92** | + | 0.85 | – | **0.85** | + | 0.10 |
| K5 | 0.15 | + | **0.93** | + | 0.84 | – | **0.90** | + | 0.21 |
| K6 | 0.68 | + | **0.96** | + | 0.82 | – | **0.82** | + | 0.07 |
| K7 | 0.83 | + | 0.90 | – | 0.84 | – | **0.91** | + | –0.42 |
| K8 | 0.59 | + | **0.96** | + | 0.00 | – | **0.91** | + | 0.32 |
| K9 | 0.68 | + | 0.95 | – | 0.85 | – | **0.85** | + | –0.41 |
| K10 | 0.54 | + | **0.93** | + | 0.85 | – | **0.85** | + | 0.10 |
| K11 | 0.54 | + | **0.93** | + | 0.85 | – | **0.85** | + | 0.10 |
| K12 | 0.02 | – | 0.94 | + | **0.84** | + | **0.91** | + | 0.68 |
| K13 | 0.08 | – | 0.94 | + | **0.85** | + | **0.89** | + | 0.69 |
| K14 | **0.82** | – | 0.94 | + | **0.83** | + | **0.87** | + | 0.87 |
| K15 | **0.98** | – | 0.95 | + | **0.82** | + | 0.78 | + | 0.88 |
| Modified AMODE | **0.95** | + | **0.94** | – | **0.86** | – | **0.97** | – | 0.93 |
| Topology constraint $n_r$ | **0.90** | + | **0.95** | – | **0.85** | – | **0.96** | – | 0.92 |

To evaluate the overall performance of each parameter across the four groups of reactivity, we introduced an average $R^2$ value, which is defined as follows:

$$\text{Average } R^2 = 1/4 \cdot \sum_i (R_i^2 \cdot (-1)^k) \tag{5}$$

where $k = 2$ ($k = 1$) if the parameter is correctly (incorrectly) capturing the expected trend. The results in Table 3 show that K1-K11 have generally low average $R^2$ values ($< 0.32$), indicating poor overall predictive performance. In contrast, parameters K12-K15, especially K14 (i.e., the so-called basicity coefficient) and K15 (i.e., the extent of depolymerization), exhibit relatively high average $R^2$ values (0.87-0.88), suggesting superior overall performance (as compared to K1-K11). The better overall performance of the basicity and the extent of depolymerization parameters may be attributed to their underlying physics: basicity is a reflection of basic oxide (network modifier dominated by ionic bonding) content over acidic oxide (network former dominated by covalent bonding) content, while the extent of depolymerization gives an average estimation of the number of non-bridging oxygens per network former. However, comparing the results with those presented in the previous sections, it is clear that both the modified AMODE and topology constraint parameters exhibit superior performance (with average $R^2$ values of 0.93 and 0.92, respectively, see Table 3) than all the compositional parameters K1-K15 commonly used in the literature and standards, including the best performing basicity and NBO/T parameters (i.e., K14 and K15, respectively). This is encouraging since, as is the case for the existing compositional parameters, the calculation of the modified AMODE (Equation 2) and topology constraint parameter (Equation 4) only depends on chemical composition.

Finally, we have also assessed the ability of parameters K1-K15 to predict other reactivity data available in refs. [3-5, 8], including the degree of reaction from NMR and/or thermodynamic modeling, compressive strength data, and TGA bound water data collected for NaOH-activated GGBSs. The levels of agreement (as evaluated using $R^2$ values for linear regression) are summarized in Table S1 of the Supplementary Material, which is generally consistent with the observations in Table 3 (i.e., the modified AMODE and topology constraint parameters exhibit better overall performance than the other parameters).

## 3.4 Application of the Modified AMODE and Topology Constraint ($n_r$) Parameters to Other Studies

Next, we evaluate the performance of the modified AMODE and topology constraint $n_r$ parameters in their ability to capture reactivity data of aluminosilicate glasses in alkaline environments from several other recent investigations. We also compare their performance with the best performing parameters in Table 3, the basicity coefficient (K14) and the commonly used extent of depolymerization parameter (K15).

### 3.4.1 Case study I

The first case study is based on experimental data from Blotevogel *et al.* [9], where the performance of four calorimetric testing protocols, and specifically their ability to predict the reactivity of GGBS, has been evaluated. Figure 6a shows the correlation between the ICC cumulative heat obtained using the four testing protocols for six GGBSs and the modified AMODE parameter calculated using Equation 2. The results show that all the ICC cumulative heat data exhibit the expected inverse correlation with the modified AMODE parameter (i.e., a higher AMODE value gives a lower ICC cumulative heat and hence a lower reactivity). It is also clear from Figure 6a that the modified R3 test protocol by Snellings and Scrivener (i.e., 24 hr ICC heat measured on a blended mixture of $Ca(OH)_2$, KOH and GGBS at 40 °C, which is slightly different from the original R3 test proposed [15, 16]) gives the strongest correlation for the modified AMODE parameter (with an $R^2$ value of 0.99), followed by the original R3 test [15, 16] ($R^2$ = 0.85).

Figure 6b shows the correlations between the ICC cumulative heat data based on the modified R3 test and three other compositional parameters, i.e., (i) topology constraint parameter $n_r$ (calculated using Equation 4 in Section 3.2), (ii) the extent of depolymerization, i.e., NBO/T (K15 in Table 2), and the basicity coefficient (K14 in Table 2). Although the positive correlations for the NBO/T and basicity coefficient and the inverse correlation for the topology constraint $n_r$ are consistent with expectations, the level of correlation (as evaluated by $R^2$ values) achieved for the commonly used NBO/T ($R^2$ value of 0.22) is much lower than that of the two other parameters ($R^2$ values of 0.62-0.84), which is then lower than that of the modified AMODE parameter shown in Figure 6a ($R^2$ value of 0.99).

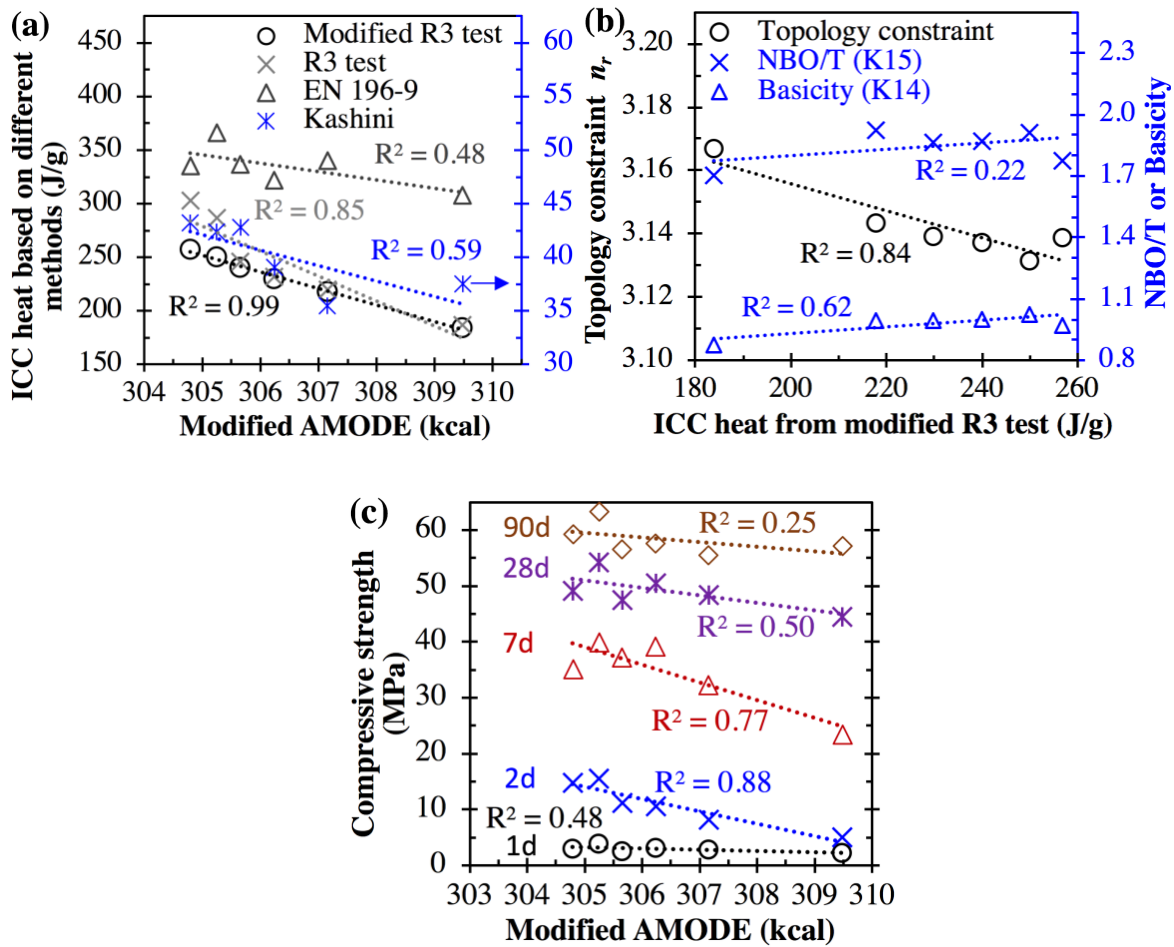

Figure 6. (a) Comparison of the modified AMODE parameter with isothermal conduction calorimetry (ICC) heat data collected using different testing methods [9], specifically the testing protocols proposed or documented by Snellings and Scrivener (modified R3 test) [14], Avet et al. [15] and Li et al. [16] (R3 test), Kashani [46], and EN 196-9 [47]. The testing protocol adopted by Snellings and Scrivener [14] has been modified from the original method, as described in detail in the original investigation [9]. (b) shows the correlation between ICC heat from the modified R3 test and three other compositional parameters, specifically, the topology constraint parameter, NBO/T, and basicity coefficient. (c) shows the correlations between the modified AMODE parameter and compressive strength data collected on a blended mixture of GGBS, clinker, gypsum, and anhydrite cured at different ages (in accordance with EN 196-1 [48]) [9].

Figure 6c shows the correlations between the modified AMODE parameter and compressive strength data collected on the six GGBSs blended with a mixture of clinker, gypsum, and anhydrite according to EN 196-1 [9]. It is seen that the compressive strength at all tested ages exhibits an inverse correlation with the modified AMODE parameter, as expected. The 2-day strength possesses the strongest correlation ($R^2$ value of 0.88), which is consistent with the original study [9], where the 2-day strength is also shown to exhibit the strongest correlation with the ICC heat data obtained using the modified R3 test. The level of correlation (for linear regression) between (i) the 2-day strength data and (ii) the compositional parameters in Table 2 and the topology constraint parameter $n_r$ are summarized in Table 4. The results show that all the parameters give the expected positive correlation, and many of these parameters (including the simplest ones, e.g., K1, K2) give comparable $R^2$ values to the modified AMODE and topology constraint parameters, while the NBO/T (K15) and basicity (K14) parameters exhibit the lowest $R^2$ values (0.15-0.50).

We note that there are additional compressive strength and ICC heat data based on the R3 test for another ten GGBSs in the referred study [9]. However, none of these well-performing parameters identified in Section 3.3 (i.e., K12-K15, modified AMODE, and $n_r$) are able to give a reasonable description of the strength and ICC data for all sixteen GGBSs ($R^2$ values < 0.3, see Table S2 and Figures S4-S5 in the Supplementary Material). It is surprising to see that the most reactive GGBS according to these compositional parameters (i.e., K12-K15, modified AMODE, and $n_r$) give the lowest ICC cumulative heat (based on the R3 test) and 2-day strength, while the supposedly least reactive GGBS exhibit the highest ICC cumulative heat and strength. The exact reason for this obvious contradiction is unclear, but we have presented some further analysis of these correlations (see Figures S4 and S5 and Table S2) and a discussion of the results in Section 6 of the Supplementary Material.

Table 4. Level of correlation between (i) the 2-day strength collected for six GGBSs in ref. [9], where four testing protocols have been used to collect ICC data, and (ii) the different compositional parameters in Table 2 as well as the modified AMODE and topology constraint parameter $n_r$. The "+" sign denotes a positive correlation, while the "−" denotes an inverse correlation. Similar to Table 3, the shaded signs indicate consistency between the predicted and expected trends.

| Parameters | $R^2$ value for linear fits | Positive (+) or inverse (−) correlation |
| --- | --- | --- |
| K1 | 0.86 | + |
| K2 | 0.86 | + |
| K3 | 0.88 | + |
| K4 | 0.89 | + |
| K5 | 0.88 | + |
| K6 | 0.88 | + |
| K7 | 0.72 | + |
| K8 | 0.78 | + |
| K9 | 0.89 | + |
| K10 | 0.88 | + |
| K11 | 0.91 | + |
| K12 | 0.74 | + |
| K13 | 0.72 | + |
| K14 | 0.50 | + |
| K15 | 0.15 | + |
| Topology constraint $n_r$ | 0.76 | − |
| Modified AMODE | 0.88 | − |

### 3.4.2 Case study II

The second case study is based on data from Durdziński et al. [31], where the reactivity of synthetic Ca-Mg-Na-aluminosilicate glasses in NaOH solutions and Portland cement has been evaluated. The equations used for calculating the modified AMODE (Equation 2) and topology constraint $n_r$ (Equation 4) parameters have been slightly modified to take into account the impact of Na, as shown in Equations 6 and 7, respectively.

$$Modified\ AMODE = \frac{218N_{Ca}+189N_{Mg}+360N_{Al}+424N_{Si}+110N_{Na}}{(N_{Ca}+N_{Mg}+N_{Al}+N_{Si}+N_{Na})} \quad (6)$$

$$n_r = \frac{3N_O+5N_{Si}+4.15N_{Al}+2N_{Ca}+2N_{Mg}+N_{Na}}{N_O+N_{Si}+N_{Al}+N_{Ca}+N_{Mg}+N_{Na}} \quad (7)$$

where $N_{Na}$ is the molar quantity of Na in each glass, and the pre-factor of $N_{Na}$ in Equation 6, i.e., 110, is estimated based on two considerations: (i) a single bond strength of 20 kcal for the Na-O bond in oxide glasses [38] and (ii) an average CN of 5.5 for Na atoms in aluminosilicate glasses (different values have been reported, e.g., 5, 5.8, and 6.1) [49]. The inclusion of Na in Equation 7 assumes that each excess Na atom beyond those required for charge-balancing the negatively charged four-fold Al atoms ($[Al(O_{1/2})_4]^{-1}$) creates one BS constraint associated with the $Na^+$-NBO bond, in accordance with ref. [37].

The log dissolution rates of these four synthetic Ca-Mg-Na-aluminosilicate glasses, along with one GGBS (which is predominantly amorphous and has a comparable particle size distribution to the synthetic glasses), are shown in Figure 7 as a function of four parameters: the modified AMODE and topology constraint $n_r$ parameters, and the commonly used NBO/T and basicity coefficient. It is clear that the log dissolution rate of these amorphous aluminosilicates is inversely correlated with the modified AMODE and topology constraint $n_r$ parameters, with a lower modified AMODE or $n_r$ value leading to a higher dissolution rate. This is consistent with the results in the previous sections, where the reactivity of aluminosilicate glasses increases with decreasing modified AMODE and $n_r$ value. As also expected, there are positive correlations observed between the experimental data and the basicity and NBO/T parameters. However, it is seen that the modified AMODE and $n_r$ parameters exhibit higher levels of correlation ($R^2 = 0.76$-$0.80$) than NBO/T ($R^2 = 0.60$) and basicity ($R^2 = 0.36$).

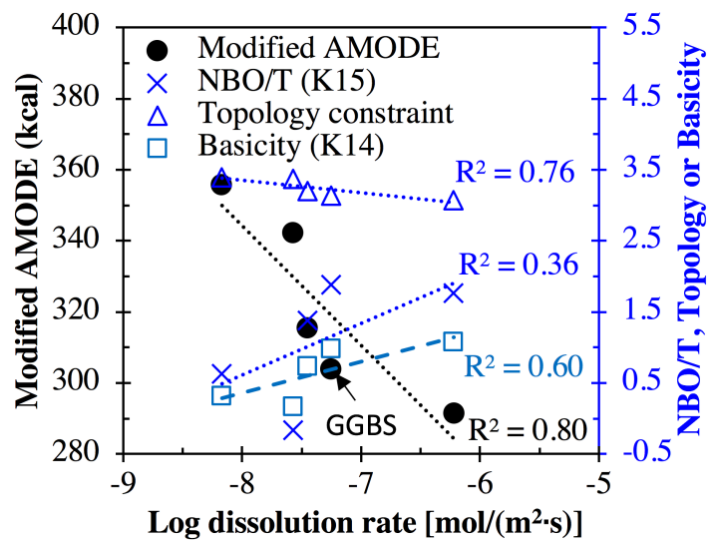

Figure 7. Comparison of the initial log dissolution rate (based on Si) from batch dissolution experiments conducted on four synthetic Ca-Mg-Na-aluminosilicate glasses and one GGBS (obtained from ref. [31]) with the four compositional parameters, i.e., modified AMODE (Equation 6), topology constraint $n_r$ (Equation 7), NBO/T (K15 in Table 2), and basicity coefficient (K14 in Table 2).

### 3.4.3 Case study III

The third case study is based on data from Nie et al. [24], where the pozzolanic reactivity of eighteen synthetic CAS and CMAS glasses in a lime-saturated solution has been investigated along with the impact of Mg substitution on Ca. Figure 8a shows the correlations between (i) the modified AMODE (Equation 2), topology constraint $n_r$ (Equation 4), NBO/T (K15 in Table 2) and basicity (K14 in Table 2) parameters for all the CAS and CMAS glasses, and (ii) their degree of reaction in lime-saturated solution after 7 days of immersion at 40 °C, as determined using $^{27}$Al NMR. The general trends in Figure 8a, i.e., the inverse correlation for the modified AMODE and topology constraint and the positive correlation for NBO/T and basicity, are consistent with expectations. However, it is clear that the levels of correlation achieved, especially for the modified AMODE and topology constraint $n_r$ parameters, are relatively low ($R^2 = 0.44$) as compared with $R^2$ values obtained in the previous sections (e.g., Figures 2, 3, 6, and 7).

Figure 8b shows the same relationship for the synthetic CAS glasses only, which exhibit higher levels of correlation (~0.75-0.86) than those seen in Figure 8a (where data for CMAS glasses have also been included). Since the original study was designed to investigate the impact of Mg substitution of Ca on the reactivity of aluminosilicate glass [24], this difference between Figures 8a and 8b suggests that the modified AMODE and topology constraint parameters (similar to NBO/T and basicity) fail to accurately capture the impact of alkaline earth cation type (Mg versus Ca) on the reactivity of aluminosilicate glasses in the saturated lime solution. This failure is expected for the topology constraint, NBO/T, and basicity parameters calculated in this article (see Equation 4 and K14-15 in Table 2), which do not distinguish the difference between Ca and Mg atoms. However, the modified AMODE parameter does distinguish the difference between Ca and Mg atoms by taking into account the differences in their CNs and the single bond strength of Ca-O and Mg-O bonds (Equation 2).

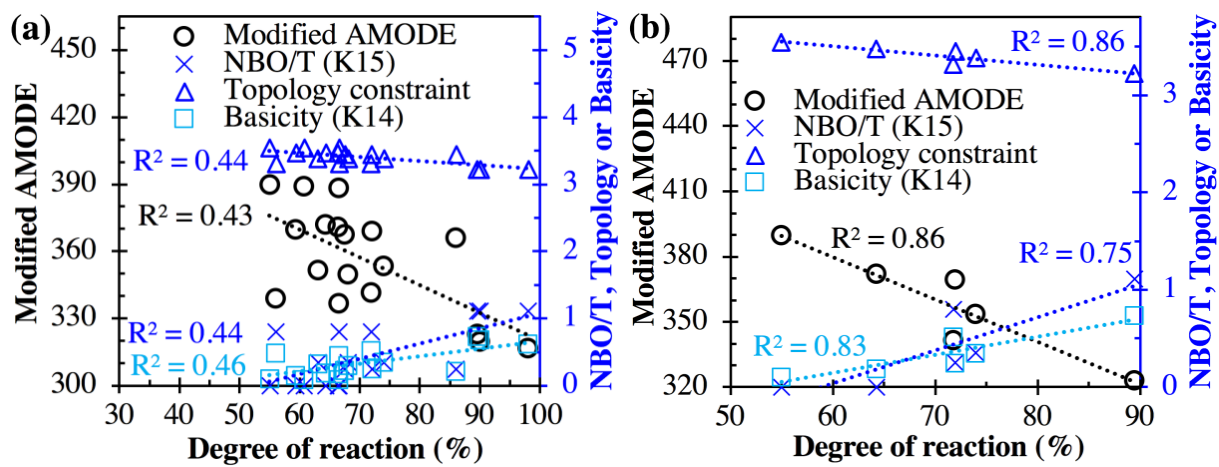

Figure 8. Comparison of (i) four compositional parameters, specifically, the modified AMODE (Equation 2), NBO/T (K15 in Table 2), topology constraint (Equation 4), and basicity (K14 in Table 2), with (ii) the degree of reaction of synthetic CAS and CMAS glasses in lime-saturated solution (reacted for 7 days at 40 °C) obtained from ref. [24]. The degree of reaction has been estimated using $^{27}$Al NMR [24]. (a) shows the data for both CAS and CMAS glasses, while (b) shows only the CAS glass data.

We further looked into the impact of the Mg/Ca ratio at fixed NBO/T and $n_r$ values on the reactivity of CAS and CMAS glasses, and the results are presented in Figure 9, which shows that both the modified AMODE parameter and the degree of reaction can vary considerably at fixed values for the NBO/T and topology constraint $n_r$ parameters. It is also seen that the expected inverse correlations between the modified AMODE and the degree of reaction are observed only at (i) high NBO/T and low $n_r$ values (Figure 9a) and (ii) low NBO/T and high $n_r$ values (Figure 9f), where increasing Mg/Ca ratio is seen to lead to a greater extent of reaction (higher reactivity). This is consistent with Equation 2 in Section 3.1, where an increase in Mg content at fixed Si and Al contents (hence an increased Mg/Ca ratio at fixed NBO/T and $n_r$ values) leads to a reduction in the modified AMODE value (as illustrated in Figure S7 of the Supplementary Material) and thus a higher reactivity. However, for intermediate NBO and $n_r$ values (Figure 9b-e), there are no obvious and consistent trends between the degree of reaction and (i) modified AMODE and (ii) Mg/Ca ratio. This suggests that the impact of Mg (as compared to Ca) on aluminosilicate glass reactivity is more complex than what the modified AMODE parameter can capture, and the reason for this complexity may be caused by several factors.

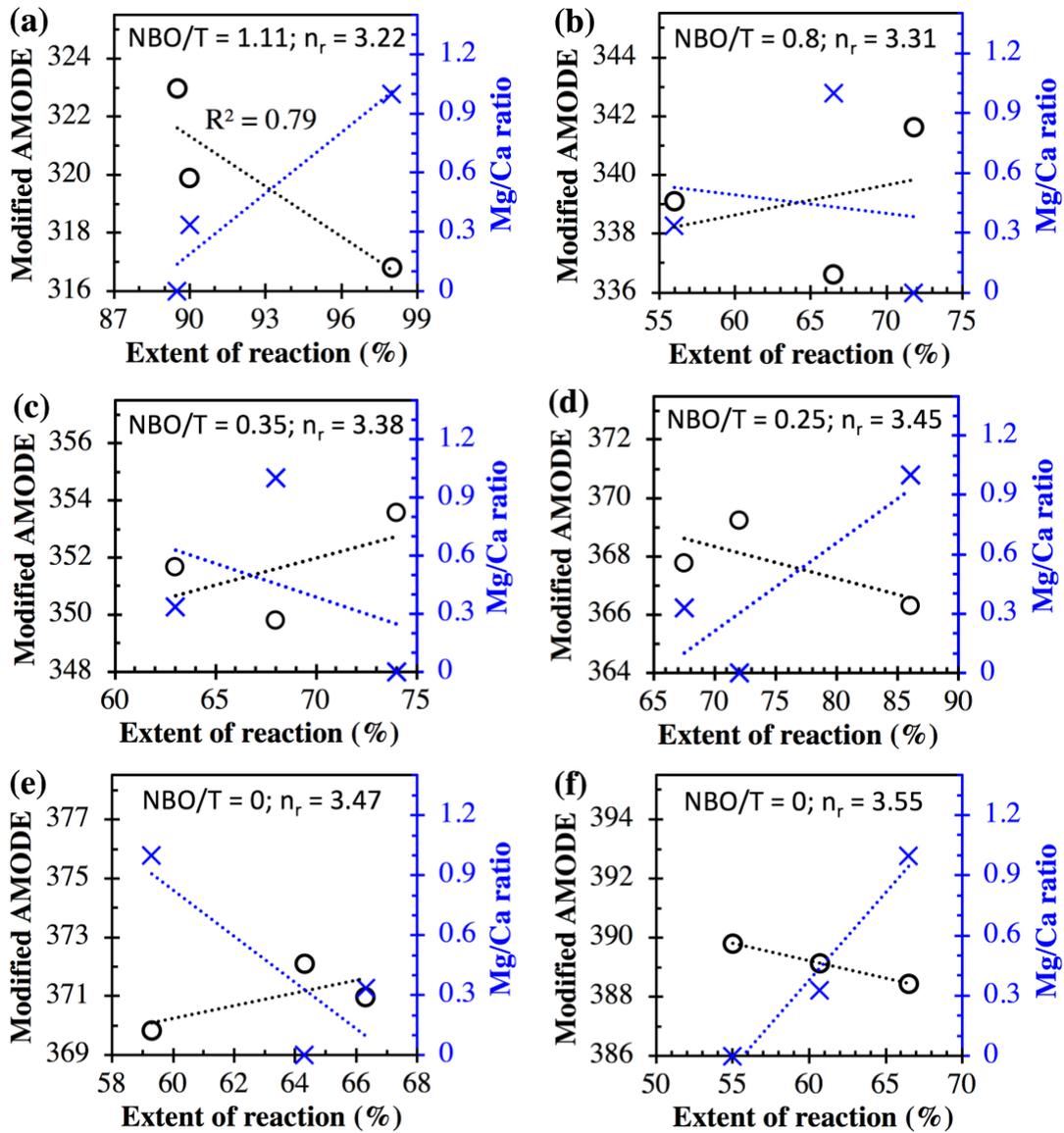

Figure 9. Comparison of (i) the modified AMODE parameter and Mg/Ca molar ratio with (ii) the degree of reaction from $^{27}$Al NMR data [24] for CAS and CMAS glasses at (a) NBO/T = 1.11 and $n_r$ = 3.22, (b) NBO/T = 0.80 and $n_r$ = 3.31, (c) NBO/T = 0.35 and $n_r$ = 3.38, (d) NBO/T = 0.25 and $n_r$ = 3.45, (e) NBO/T = 0.00 and $n_r$ = 3.47, and (f) NBO/T = 0.00 and $n_r$ = 3.55. The $n_r$ value was calculated using Equation 4 in Section 3.2, while the NBO/T value has been estimated using equation K15 in Table 2, where the molar contents of Na$_2$O, K$_2$O, MnO, FeO, and Fe$_2$O$_3$ in equation K15 are taken to be zero.

First, the single bond strength values for Ca-O and Mg-O bonds (i.e., 32 and 37 kcal, respectively) adopted for the calculation of the modified AMODE parameter are for six-fold Ca and Mg atoms.

According to our previous MD simulations [13], both Ca and Mg atoms have a distribution of CNs, with predominately seven- and six-fold coordination for Ca atoms and five-fold for Mg atoms. Hence, the actual single bond strength values may differ slightly from those used in this investigation (in Equation 2). Second, the modified AMODE parameter uses the average CNs (6.81 for Ca atom and 5.10 for Mg atom, as given in Table 1) based on MD simulations of CMAS and CAS glasses with compositions different from those in ref. [24]. Hence, deviations of these average CN values from the actual Ca and Mg CNs in each CMAS and CAS glass in Figure 8 are expected. Errors caused by these deviations could be reduced by performing MD simulations and calculating the corresponding CN values for each CAS and CMAS glass composition in Figure 8, enabling the calculation of the original AMODE parameter introduced in ref. [13]. But this need for additional MD simulations and/or experiments goes against the original reasoning behind the motivation for a compositional parameter in the first place, i.e., the ability to predict reactivity based on chemical composition alone. Nevertheless, further refinement of the modified AMODE parameter, taking into account the more nuanced effects of Ca and Mg, is suggested as a future work direction. For example, the development of machine learning models based on high throughput MD simulations is a promising direction for predicting Ca and Mg CNs as a function of C(M)AS glass compositions, although one needs to consider the accuracy of force field MD in capturing Ca and Mg CNs.

The third possible reason why the modified AMODE parameter cannot accurately capture the impact of Mg (as compared to Ca) on aluminosilicate glass reactivity in alkaline environments relates to the atomic structural complexity of these glasses (more complex than silicate minerals). In alkaline earth metal silicate mineral dissolution experiments, the Ca dissolution rate is generally several orders higher than that of Mg at the same level of NBO/T (e.g., dicalcium silicate vs. dimagnesium silicate with an NBO/T of 4) and dissolution conditions [32]. This observation for silicate mineral dissolution is consistent with our previous DFT calculations on a CMAS glass, which show that the electron density of states around Ca atoms in the valence band near the Fermi level is much higher than that around Mg atoms, suggesting a higher reactivity of the former (i.e., Ca atoms). On the other hand, Mg atoms are seen to promote the formation of free oxygen sites (as compared to Ca atoms) in CMAS glasses [39, 50], which do not exist in common alkaline earth metal silicate minerals. These free oxygen sites (defined as oxygen atoms only connected to network modifiers, e.g., Ca and Mg) are the most reactive oxygen species in CMAS glasses, as evidenced by the electron density of state calculations for different oxygen species in ref. [39]. Hence, this dual impact of Mg on CAS glass reactivity (not captured by the modified AMODE parameter) may have contributed to the seemingly chaotic trends seen in Figures 9b-e and hence the relatively low $R^2$ value seen in Figure 8a.

In addition, all the compositional parameters (including the modified AMODE, topology constraints, and the K parameters shown in Table 2) assume congruent dissolution of all oxide components, which may not be the case in experiments. In case study III, the degree of reaction was based on the amount of reacted Al, obtained using $^{27}$Al NMR data and the assumption of congruent dissolution. Although there is evidence to show close to a congruent dissolution of Si and Al atoms from CAS glasses in high pH environments [51], there are some levels of incongruency (i) between network formers and network modifiers and (ii) between Ca and Mg cations, as illustrated by Figure S8 in the Supplementary Material, based on alkaline earth aluminosilicate glass dissolution data reported in ref. [51]. Hence, any deviations from congruent dissolution may have also contributed to the relatively low $R^2$ value seen in Figure 8a. Although this factor could have contributed to the observed correlations in the previous sections, the extent of incongruency for glass dissolution (and hence its impact) may vary considerably depending on the experimental conditions, including the pH, temperature, and solution chemistry [52, 53]. Recently, the potential impact of incongruent dissolution on the ability of the original AMODE parameter to predict the relative reactivity of complex natural volcanic glasses in an alkaline environment has also been discussed [36].

Finally, as mentioned in Sections 3.1-3.2, the experimental data in Figures 2 and 3 has been normalized by the surface area of each GGBS or CAS glass following ref. [13]. According to case study I [9], the ICC cumulative heat release in a blended mixture of Ca(OH)$_2$, KOH, and GGBS is linearly and positively correlated with the specific surface area (from BET measurement) of the GGBS particles (with $R^2$ values of 0.97-1.00 for linear regression). Although caution has been taken in case study III to select glass particles of the same size range (20-40 $\mu$m), the reactivity data has not been normalized by the surface area of the C(M)AS glasses, and thus there could be differences in the specific surface area of the synthetic glasses. These differences might have also contributed to the relatively low $R^2$ value values seen in Figure 8 and the chaotic trends in Figure 9b-e.

## 4 Broader Impact and Limitations

### 4.1 Broader impact

Despite the varying levels of performance of the two compositional parameters introduced here (the modified AMODE and topology constraint parameters) for capturing the different reactivity data of amorphous aluminosilicates relevant to cement and concrete, the results presented here are encouraging, especially considering the strong correlation seen between the modified AMODE parameter and the modified R3 test data in Figure 6 ($R^2$ value of 0.99 for linear regression). The modified R3 test has been widely used in the cement and concrete community to gauge the reactivity of SCMs and hence their suitability to be used for concrete production. While the experimental reactivity tests (e.g., the modified R3 test) require at least 24 hours and access to a testing instrument, the modified AMODE parameter can be calculated from the chemical composition of SCMs (often measured by industry) in a matter of minutes. Hence, there could be some economic benefits and time savings by developing the modified AMODE parameter into a reliable fast-screen method for SCM reactivity (e.g., screening out some SCMs for further experimental validation as opposed to performing reactivity experiments on all SCMs).

Furthermore, it is conceivable that the method of deriving these physics-based compositional parameters can be extended to investigate the reactivity of (i) more complex aluminosilicate glasses containing other oxide components, e.g., FeO, $K_2O$, $Na_2O$, MnO, and $TiO_2$, and (ii) other types of glasses, e.g., borosilicate glasses for nuclear waste encapsulation and bioactive glasses for bone regeneration and drug delivery applications. A demonstration has been given in case study II, where Equations 2 and 4 have been extended to cover CMAS glasses containing $Na_2O$ (Equations 6 and 7 for calculating the modified AMODE and topology constraint parameters, respectively). Similarly, the method of derivation may also be extended to study mineral or clay dissolution in the geochemistry community, although the associated limitations (as have been discussed in Section 3.4.3 and will be summarized in Section 4.2) need to be carefully considered.

## 4.2 Limitations

Despite the simplicity of applying these two physics-based compositional parameters to predict CAS and CMAS glass reactivity in alkaline environments and their generally superior performance in comparison to other compositional parameters in Table 2, several limitations warrant discussion. The reactivity of amorphous aluminosilicates in alkaline environments is highly complex, as illustrated by case study III on CMAS glasses. In addition to the composition and structure of CAS and CMAS glasses, many other factors can dramatically impact their reactivity in blended Portland cements or AAM systems, including solution chemistry, curing conditions, particle size distribution, and level of amorphicity [1, 7, 9-12]. It is also important to keep in mind that many SCMs used in blended cements, e.g., coal-derived fly ash, are heterogeneous and more complex in composition and mineralogy than those presented in this investigation, which are either pure synthetic glasses or GGBSs with a high level of amorphicity. Furthermore, the potential phase segregation in the glassy phases of the SCMs, as has been shown to be the case for fly ash [31, 54], could have a dramatic impact on their reactivity in alkaline environments. The presence of mineral phases and the segregation of glassy phases represent significant challenges for the conversion of compositional parameters (including the modified AMODE and topology constraint parameters introduced here) into fast-screen methods.

In addition to the limitations outlined above regarding SCM complexity, there are a number of limitations associated with the estimation of these physics-based compositional parameters that need to be considered, as alluded to in Section 4.3:

- The assumption of congruent dissolution for all the compositional parameters, including the modified AMODE and topology constraint parameters, may not apply. The extent of deviation from congruent dissolution can vary considerably depending on the glass composition and dissolution conditions (e.g., solution chemistry, pH, and temperature) [51-53, 55]. Hence, in the future, it would be worth exploring how these compositional parameters (e.g., modified AMODE and topology constraint parameters) should be modified to take into account the impact of incongruent dissolution.

- The deviation of the single metal-oxygen bond strength values adopted for calculating the modified AMODE parameter from the actual values, especially concerning the Al-O bond strength in five-fold coordination.
- Difference between the average CNs of the Al, Ca, and Mg atoms used for calculating the modified AMODE parameter in Equation 2 and the actual CN values in each aluminosilicate glass. As shown in Table 1, 2-11% differences are observed between the average CNs values based on all the eighteen CAS and CMAS glasses in Table 1 and the CNs of individual glasses from MD simulation results.
- The CNs of some glass compositions may not be accurately captured by the force field MD simulations, as discussed in our previous investigation [13], where the MD simulations are seen to significantly underestimate the proportion of five-fold Al atoms in highly peraluminous CAS glasses ($Al_2O_3 > CaO$).

## 5  Conclusions

The reactivity of aluminosilicate glasses in alkaline environments is critical to many important industrial applications, including blended Portland cements and alkali-activated materials. In this investigation, two physics-based compositional parameters have been derived for rapid evaluation of the relative reactivity of $CaO-Al_2O_3-SiO_2$ (CAS) and $CaO-MgO-Al_2O_3-SiO_2$ (CMAS) glasses in alkaline environments. The first parameter, the modified average metal oxide dissociation energy (AMODE), has been derived from recent molecular dynamics (MD) simulation results [13], while the second parameter, the average topology constraint per atom, $n_r$, has been derived from topological constraint theory (TCT). Both parameters are seen to give an accurate description of different reactivity data (e.g., degree of reaction) collected on the eighteen CAS and CMAS glasses from four literature investigations [3-5, 8], with $R^2$ values (generally higher than 0.90 for linear regressions with different reactivity data) comparable to that of the original AMODE parameter introduced in our previous study [13]. The predictive performance of the modified AMODE and topology constraint ($n_r$) parameters has been compared with that of fifteen other compositional parameters obtained from the literature, where the results show that the two physics-based compositional parameters are generally the most accurate for predicting the relative reactivity of CAS and CMAS glasses in alkaline environments.

To further evaluate the performance of the two physics-based compositional parameters in comparison with the best performing parameters identified from the literature, i.e., the basicity coefficient and commonly used extent of depolymerization parameter (NBO/T), we have extended the analysis to three recent investigations that have explored the reactivity of GGBSs [9], synthetic Na-containing CMAS glasses [31], and synthetic CAS and CMAS glasses [24] in different alkaline environments. The results show that both the modified AMODE and topology constraint ($n_r$) parameters generally outperform the basicity coefficient and NBO/T parameters, with $R^2$ values of > 0.80 for linear regression with different reactivity data. One exception is the $^{27}$Al NMR-derived reactivity data for CMAS glasses from Nie et al. [24], where the levels of correlation with all four parameters (i.e., modified AMODE, topology constraint, basicity coefficient and NBO/T) are relatively low ($R^2$ = 0.43-0.46), although the general trends are correctly captured. This has been attributed to the complex impact of Mg and Ca cations on aluminosilicate glass reactivity, along with several limitations associated with the use of compositional parameters, including the assumption of congruent dissolution, which may not be representative of actual experiments. Nevertheless, all analyses have shown that the modified AMODE and topology constraint parameters generally outperform the other compositional parameters available in the existing literature. Given that both parameters are physics-based (e.g., containing structural information) and can be calculated directly from chemical compositions (without the need for experiments or simulations), they hold great promise for widespread application.

## 6  Supplementary Material

Correlation between the modified AMODE parameter and the additional experimental data; Correlation between the topology constraint parameter $n_r$ and the additional experimental data; Impact of minor oxides on parameters K8-K11; Correlation between reactivity data in Figure 5 and K1, K2, K4, K6-K12 and K14 parameters; Correlation between K1-K15 and the additional experimental data; Correlations with the compressive strength and ICC cumulative heat data in case study I; Impact of the Mg/Ca ratio on the modified AMODE parameter; Evaluation of the level of congruency for CAS and CMAS glass dissolution in alkaline environments.

## 7 Acknowledgments

This material is based on work supported by ARPA-E under Grant No. 1953-1567.